\documentclass[pdflatex]{svmult}

\usepackage{makeidx}     
\usepackage{graphicx}    
\usepackage{multicol}    

\makeindex             

\newcommand{\EEq}[1]{Equation~(\ref{#1})}
\newcommand{\Eq}[1]{equation~(\ref{#1})}
\newcommand{\eq}[1]{(\ref{#1})}  
\newcommand{\Fig}[1]{Fig.~\ref{#1}}
\newcommand{\nab}{\mbox{\boldmath $\nabla$} {}}
\newcommand{\meanC}{\overline{C}}
\newcommand{\meanFF}{\overline{\mbox{\boldmath ${\mathcal F}$}} {}}
\newcommand{\uu}{\mbox{\boldmath $u$} {}}
\def\onethird{{\textstyle{1\over3}}} 
\newcommand{\bra}[1]{\langle #1\rangle}
\newcommand{\yprl}[4]{ (#1) #4. Phys.\ Rev.\ Lett.\ #2:#3}
\newcommand{\ypf}[5]{ (#1) #5. Phys.\ Fluids #2:#3--#4}
\newcommand{\yjour}[6]{ (#1) #6. #2. #3:#4--#5}

\begin{document}

\title*{Passive scalar diffusion as a damped wave}
\author{Axel Brandenburg\inst{1}
\and Petri J.\ K\"apyl\"a\inst{2,3}
\and Amjed Mohammed\inst{4}}
\institute{NORDITA, Blegdamsvej 17, 2100 Copenhagen \O, Denmark,
\texttt{brandenb@nordita.dk}
\and Kiepenheuer-Institut f\"ur Sonnenphysik, Sch\"oneckstr.\ 6, 79104 Freiburg, Germany
\and Department of Physical Sciences, Astronomy Division, P.O. Box 3000, FIN-90014 University of Oulu, Finland,
\texttt{petri.kapyla@oulu.fi}
\and Physics Department, Oldenburg University, 26111 Oldenburg, Germany
\texttt{amjed@mail.uni-oldenburg.de}}
%
%
\maketitle

Three-dimensional turbulence simulations are used to show that the
turbulent root mean square velocity is an upper bound of the speed of
turbulent diffusion.
There is a close analogy to magnetic diffusion where the maximum diffusion
speed is the speed of light.
Mathematically, this is caused by the inclusion of the Faraday
displacement current which ensures that causality is obeyed.
In turbulent diffusion, a term similar to the displacement current
emerges quite naturally when the minimal tau approximation is used.
Simulations confirm the presence of such a term and give a quantitative
measure of its relative importance.

\section{Introduction}

Since the seminal paper of Prandtl (1925), turbulent diffusion has always
been an important application of turbulence theory.
By analogy with the kinetic theory of heat conduction, the turbulent exchange
of fluid elements leads to an enhanced flux, $\meanFF$, of a
passive scalar concentration that is proportional to the negative mean
concentration gradient,
\begin{equation}
\meanFF=-\kappa_{\rm t}\nab\meanC\quad\mbox{(Fickian diffusion)},
\label{Fickian}
\end{equation}
where $\kappa_{\rm t}={1\over3}u_{\rm rms}\ell_{\rm cor}$ is a turbulent diffusion
coefficient, $u_{\rm rms}$ is the turbulent rms velocity, and $\ell_{\rm cor}$
is the correlation length.
\EEq{Fickian} leads to a closed equation for the evolution of the mean
concentration,
$\meanC$,
\begin{equation}
{\partial\meanC\over\partial t}=\kappa_{\rm t}\nabla^2\meanC.
\label{elliptic}
\end{equation}
This is an elliptic equation, which implies that signal propagation is
instantaneous and hence causality violating.
For example, if the initial $\meanC$ profile is a $\delta$-function,
it will be a gaussian at the next instant, but gaussians have already
infinite support.

The above formalism usually emerges when one considers the microphysics of
the turbulent flux in the form $\meanFF=\overline{\uu\int\dot{c}\,{\rm d}t}$,
where $\dot{c}\approx-\uu\cdot\nab\meanC$ is the linear approximation to the
evolution equation for the fluctuating component of the concentration.
Recently, Blackman \& Field (2003) proposed that one should instead
consider the expression
\begin{equation}
\partial\meanFF/\partial t=\overline{\dot{\uu}c}+\overline{\uu\dot{c}}.
\label{dmeanFFdt}
\end{equation}
On the right hand side, the nonlinear terms in the two evolution equations
for $\uu$ and $c$ are {\it not} omitted; they lead to triple correlations
which are {\it assumed} to be proportional to $-\meanFF/\tau$, where
$\tau$ is some relaxation time.
Furthermore, there is a priori no reason to omit the time derivative on
the left hand side of \Eq{dmeanFFdt}.
It is this term which leads to the emergence of an extra time derivative
(i.e.\ a `turbulent displacement flux')
in the modified `non-Fickian' diffusion law,
\begin{equation}
\meanFF=-\kappa_{\rm t}\nab\meanC-\tau{\partial\meanFF\over\partial t}
\quad\mbox{(non-Fickian)}.
\label{nonFickian}
\end{equation}
This turns the elliptic equation \eq{elliptic} into a damped wave equation,
\begin{equation}
{\partial^2\meanC\over\partial t^2}
+{1\over\tau}{\partial\meanC\over\partial t}
=\onethird u_{\rm rms}^2\nabla^2\meanC.
\label{nonFickian_evol}
\end{equation}
The maximum wave speed is obviously $u_{\rm rms}/\sqrt{3}$.
Note also that, after multiplication with $\tau$, the coefficient on the
right hand side becomes $\onethird\tau u_{\rm rms}^2=\kappa_{\rm t}$,
and the second time derivative on the left hand side becomes unimportant
in the limit $\tau\to0$, or when the physical time scales are long compared
with $\tau$.

\section{Validity of turbulent displacement flux and value of $\tau$}
\label{Sinitialflux}

\begin{figure}[t!]\begin{center}
\includegraphics[width=.6\textwidth]{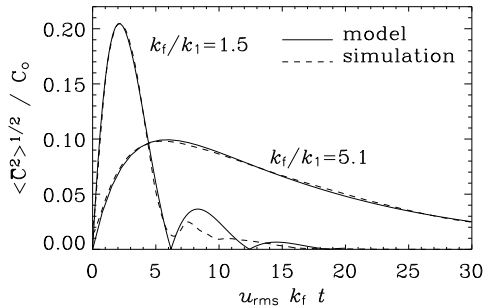}
\end{center}\caption[]{
Passive scalar amplitude, $\bra{\meanC^2}^{1/2}$,
versus time (normalized by $u_{\rm rms}k_{\rm f}$)
for two different values of $k_{\rm f}/k_1$.
The simulations have $256^3$ meshpoints.
The results are compared with solutions to the
non-Fickian diffusion model.
}\label{plncc_comp}\end{figure}

A particularly obvious way of demonstrating the presence of the second
time derivative is by considering a numerical experiment where $\meanC=0$
initially.
\EEq{elliptic} would predict that then $\meanC=0$ at all times.
But, according to the alternative formulation \eq{nonFickian_evol},
this need not be true if initially $\partial\meanC/\partial t\neq0$.
In practice, this can be achieved by arranging the initial fluctuations
of $c$ such that they correlate with $u_z$.
Of course, such highly correlated arrangement will soon disappear and
hence there will be no turbulent flux in the long time limit.
Nevertheless, at early times, $\bra{\meanC^2}^{1/2}$ (a
measure of the passive scalar amplitude) rises from zero to a finite
value; see \Fig{plncc_comp}.

\begin{figure}[t!]\begin{center}
\includegraphics[width=.72\textwidth]{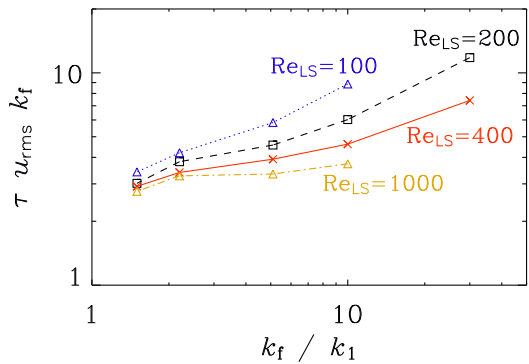}
\end{center}\caption[]{
Strouhal number as a function of $k_{\rm f}/k_1$ for different
values of $\mbox{Re}_{\rm LS}$, i.e.\ the large scale Reynolds number.
The resolution varies between $64^3$ meshpoints ($\mbox{Re}_{\rm LS}=100$)
and $512^3$ meshpoints ($\mbox{Re}_{\rm LS}=1000$).
}\label{pstrouhal_all_col}\end{figure}

Closer inspection of \Fig{plncc_comp} reveals that when the wavenumber
of the forcing is sufficiently small (i.e.\ the size of the turbulent
eddies is comparable to the box size), $\bra{\meanC^2}^{1/2}$ approaches
zero in an oscillatory fashion.
This remarkable result can only be explained by the presence of
the second time derivative term giving rise to wave-like behavior.
This shows that the presence of the new term is actually justified.
Comparison with model calculations shows that the non-dimensional
measure of $\tau$, $\mbox{St}\equiv\tau u_{\rm rms} k_{\rm f}$,
must be around 3.
(In mean-field theory this number is usually called Strouhal number.)
This rules out the validity of the quasilinear (first order smoothing)
approximation which would only be valid for $\mbox{St}\to0$.

\begin{figure}[t!]\begin{center}
\includegraphics[width=.6\textwidth]{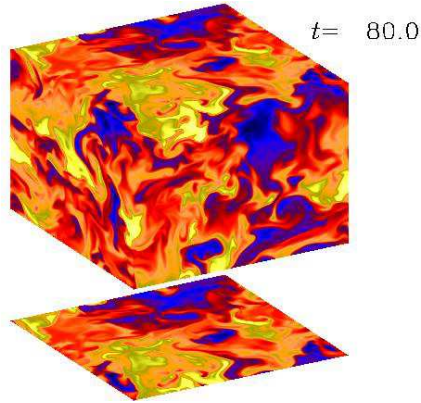}
\end{center}\caption[]{
Visualizations of $C$ on the periphery of the simulation domain
at a time when the simulation has reached a statistically steady state.
$k_{\rm f}/k_1=1.5$, $\mbox{Re}_{\rm LS}=400$.
}\label{nolog256a}\end{figure}

Next, we consider an experiment to establish directly the value of St.
We do this by imposing a passive scalar gradient, which leads to a steady
state, and measuring the resulting turbulent passive scalar flux.
By comparing double and triple moments we can measure St quite accurately
without invoking a fitting procedure as in the previous experiment.
The result is shown in \Fig{pstrouhal_all_col} and confirms that
$\mbox{St}\approx3$ in the limit of small forcing wavenumber, $k_{\rm f}$.
The details can be found in Brandenburg et al.\ (2004).
A Visualization of $C$ on the periphery of the simulation domain
is shown in \Fig{nolog256a} for $k_{\rm f}=1.5$.
Note the combination of large patches (scale $\sim1/k_{\rm f}$)
together with thin filamentary structures.

Finally, we should note that \Eq{dmeanFFdt} in the passive scalar
problem was originally motivated by a corresponding expression for the
electromotive force in dynamo theory, where the $\dot{\uu}$ terms leads to
the crucial nonlinearity of the $\alpha$-effect (Blackman \& Field 2002).

\printindex
\end{document}